\documentclass[prb,twocolumn,aps,epsfig]{revtex4}
\newcommand{\psbild}[1]{#1}
\usepackage{epsfig}
\newcommand{\Neel}{N\'{e}el }
\newcommand{\be}{\begin{equation}}
\newcommand{\ee}{\end{equation}}
\newcommand{\ben}{\begin{eqnarray}}
\newcommand{\een}{\end{eqnarray}}
\newcommand{\ra}{\rangle}
\newcommand{\la}{\langle}
\begin{document}
\title
{
Coupled Cluster Treatment of the Shastry-Sutherland Antiferromagnet
}
\author
{
R. Darradi$^a$, J. Richter$^a$, and D. J. J. Farnell$^b$
}
\affiliation
{
$^a$Institut f\"ur Theoretische Physik, Otto-von-Guericke Universit\"at
Magdeburg, \\
P.O.B. 4120, 39016 Magdeburg, Germany \\   
$^b$Unit Of Ophthalmology, Department of Medicine,
University Clinical Departments, Daulby Street,
University of Liverpool, Liverpool L69 3GA, United Kingdom\\
}

\date{\today}
\begin{abstract}
We consider the zero-temperature properties of the 
spin-half two-dimensional Shastry-Sutherland antiferromagnet
by using  a high-order coupled cluster method (CCM) 
treatment. We find that this model demonstrates various ground-state phases
(N\'{e}el, magnetically disordered, orthogonal dimer), 
and we make  predictions for the positions of the 
phase transition points. In particular, we find 
that orthogonal-dimer state becomes the ground state
at ${J}^{d}_2/J_1 \sim 1.477$. For the critical 
point $J_2^{c}/J_1$ where the semi-classical N\'eel order 
disappears we obtain a significantly lower value than 
$J_2^{d}/J_1$, namely, ${J}^{c}_2/J_1$ in the range  
$[1.14, 1.39]$. We therefore conclude that an intermediate 
phase exists between the \Neel and the dimer phases. 
An analysis of the energy of a competing spiral phase 
yields clear evidence that the spiral phase does not become the
ground state for any value of $J_2$.
The intermediate phase is therefore 
magnetically disordered but may exhibit 
plaquette or columnar dimer ordering. 
\end{abstract}


\maketitle

\section{Introduction}
The study of two-dimensional (2D) quantum magnetism has attracted much
experimental and theoretical attention over many years.
In 2D antiferromagnets at zero temperature
the competition between interactions and quantum fluctuations is
well balanced and one sees magnetic long-range order (LRO) as
well as magnetic disorder, dependent on details of the lattice
\cite{lhuillier01,moessner01,lhuillier03,Richter04}.  
 In particular, frustration may lead to the breakdown of semi-classical 
N\'eel LRO in  2D  quantum antiferromagnets.
Much research activity in this area
has been focused on  frustrated spin-half Heisenberg antiferromagnets
on the square lattice, such as the
$J_1$-$J_2$ model with competing antiferromagnetic nearest-neighbor $J_1$
and next-nearest-neighbor $J_2$ bonds 
(see, e.g., Refs. \cite{squa_ref7,squa_ref8,squa_ref9,
squa_ref10,Sushkov01,capriotti01a,capriotti01} 
and references therein), where a quantum paramagnetic phase near $J_2 \sim
0.5J_1$ is observed the nature of which is still under discussion.
Another canonical model is
the Shastry-Sutherland antiferromagnet 
introduced in the eighties \cite{Shastry},
which has special arrangement of frustrating next-nearest-neighbor 
$J_2$ bonds  on the square lattice, cf. Fig.\ref{fig1}.
We note that for bonds of equal strength, i.e., $J_1=J_2$, 
the Shastry-Sutherland model is equivalent to a Heisenberg model 
on one of
the eleven uniform Archimedean lattices\cite{Richter04}.
Although the initial motivation to study 
this special frustrated square-lattice 
antiferromagnet is related to the existence 
of a simple singlet-product
eigenstate (which becomes the ground state (GS) for strong frustration), the
renewed interest in the last years was stimulated by the discovering of
the new quantum phase in  $\mathrm{SrCu}(\mathrm{BO}_3)_2$
\cite{kageyama99,Miyahara} which can be understood in terms of the
Shastry-Sutherland model. 
Although the GS of this model
in the limit of small frustration $J_2$ and large $J_2$ is
well understood, the GS phase at moderate $J_2$ 
 is still a matter of
discussion.

In this paper, we study the GS phase
diagram for spin half Shastry-Sutherland model using a 
high-order coupled cluster treatment. 
The coupled cluster method (CCM) has previously been
applied to various  quantum spin systems with much
success \cite{rog_her90,squa_ref10,krueger00,krueger01,bishop94,re,
zeng98,rachid04,bishop00,Farnell04}.
 We mention 
that one particular advantage of this approach 
consists in applicability to strongly frustrated quantum spin
systems in any dimension, 
where some other methods, such as, e.g., the quantum Monte Carlo method  
fail.

\maketitle

\section{The Model}
\begin{figure}[ht]
\epsfig{file=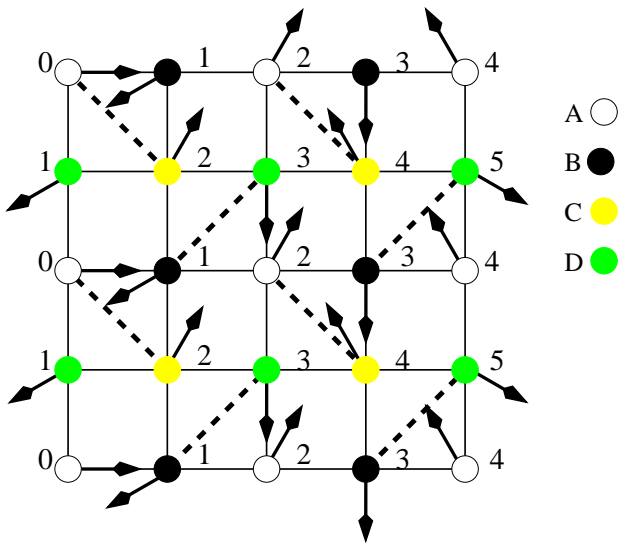,scale=0.66,angle=0}
\caption{\label{fig1} Illustration of the classical spiral state for 
Shastry-Sutherland model
of Eq.~(\ref{eq1}), with nearest-neighbor bonds $J_1$ (solid lines) and next-nearest-neighbor bonds
$J_2$ (dashed lines). The spin orientations at A, C and B, D lattice sites are defined by the angles
$\theta=n\phi$ and $\theta=n\phi+\pi$, respectively, where $n=0, 1, 2,\dots,$ and $\phi$
is the characteristic angle of the spiral state. The state is shown for $\phi=\pi/6$ and
$n=0, 1\dots 5$.}
\end{figure}

The Shastry-Sutherland model is a 
spin-$\frac{1}{2}$ Heisenberg model on a square lattice
with antiferromagnetic nearest-neighbor bonds $J_1$ 
and with one antiferromagnetic diagonal 
bond $J_2$ in each second square (see 
Fig.\ref{fig1}). It is described by the Hamiltonian
\begin{equation}
\label{eq1}
H = J_1\sum_{\langle i,j \rangle}s_i \cdot s_j + J_2\sum_{\{i, k\}}s_i \cdot s_k ,
\end{equation}
where the operators $s_i$ represent spin-half operators, i.e., $s_i=s(s+1)$
with $s=1/2$.   
The sums over $\langle i,j \rangle$ and $\{i, k\}$ run over all nearest-neighbor bonds
and
over some of the next-nearest-neighbor bonds according to the pattern 
shown in Fig. \ref{fig1}.
Due to the special arrangement of the $J_2$ bonds the unit cell contains
four sites. Therefore it is convenient to split
the square lattice into four equivalent sublattices A, C, B and D 
as shown in Fig.(\ref{fig1}). It what follows we  
set $J_1 = 1$ and consider $J_2 > 0$ as the parameter of the model. 

The classical (i.e., $s \to \infty$) 
GS of the Shastry-Sutherland model 
is the collinear N\'eel state
for 
$J_2/J_1 \le 1$, but a 
noncollinear spiral state  for $J_2/J_1 > 1$ (see Fig.\ref{fig1} and
Refs. \onlinecite{Mila,Weihong02}) 
with a characteristic 
pitch angle $\phi$ given by
\begin{eqnarray}
\label{eq2}
\phi = \left\{ \begin{array}{ll}
0 & \quad J_2 \le J_1\\
\pi - \arccos (-J_1 / J_2) & \quad J_2 > J_1
\end{array} \right.
\end{eqnarray}
We note that for  $\phi=0$ the spiral state becomes the collinear \Neel state
classically. The transition from the collinear \Neel to
noncollinear spiral state is of second order and takes place at 
 $J_2/J_1=1$. We note further 
that there are only two different angles between interacting spins, 
namely, 
$\phi+\pi$ for the $J_1$ couplings and $-2\phi$ for the $J_2$ couplings.

The quantum $s=1/2$ version of the  model has been treated 
previously by various methods like 
Schwinger boson mean-field theory \cite{Mila}, exact diagonalization 
\cite{Miyahara,Lauchli},
series expansions \cite{Weihong,Weihong02,Muller,Kawakami}, renormalization
group\cite{Hajii} 
and also by a gauge-theoretical approach
\cite{Subir}. A recent review can be found in Ref. \onlinecite{Miyahara03}. 
From these studies one knows that  
for small $J_2 \le J_1$ the physics of the quantum model is 
similar to that of 
the classical model, i.e., we have semi-classical \Neel order.
Furthermore, one knows already from  the early work of Shastry and
Sutherland \cite{Shastry} that for large  $J_2$ the
quantum GS is a
 rotationally invariant 
product state of local pair singlets (so-called orthogonal-dimer state) 
$|\Psi\rangle_{dimer}  = 
\prod_{\{i,j\}_{J_2} } [|\uparrow_{i} \rangle|\downarrow_j \rangle
-|\downarrow_{i} \rangle|\uparrow_j \rangle]/\sqrt{2}\;$, 
where 
$i$ and $j$ correspond to those sites which cover the $J_2$ bonds. 
The energy per site of this  orthogonal-dimer state
is $E_{dimer}/N= -3J_2/8$. It  becomes the GS at around 
$J_2^c \approx (1.44 \dots 1.49) J_1$ (see Table 2 in Ref.
\onlinecite{Miyahara03}). 
Note that such an  orthogonal-dimer state can be observed also in
corresponding one-dimensional and three-dimensional models
\cite{ivanov,ueda99,koga02}.
The nature of the transition between the semi-classical \Neel state and
the orthogonal-dimer phase is still a matter of
controversial discussion.
In the region $1.2 J_1 \lesssim J_2 \lesssim 1.45 J_1 $ the main 
question is
whether the system has an intermediate phase. 
A direct transition between the \Neel phase and the orthogonal-dimer phase
is favored in Refs. \onlinecite{Miyahara,Weihong,Muller,Hajii}, whereas
in Refs. \onlinecite{Mila,Subir,Weihong02,Lauchli} the existence of an
intermediate phase is found. However, concerning the nature of this 
intermediate phase controversial results are reported,  
as candidates for the intermediate phase are quantum
spiral phases \cite{Mila,Subir} or a plaquette or columnar 
singlet phases \cite{Weihong02,Lauchli} discussed. 

To contribute to the solution of this open problem  
the CCM  is an appropriate method, since it is one of the methods  
which can deal 
with spiral phases in  quantum spin models.
\cite{krueger00,krueger01,bursill}

\section{The Coupled Cluster Method}\label{ccm}
The CCM formalism is now briefly considered, although the 
interested reader is referred to Refs.
\onlinecite{re,zeng98,krueger00,bishop00,Farnell04} for
further details. 
The starting point for the CCM calculation is the choice of 
a normalized reference or model state 
$|\Phi\rangle$. For spin systems, 
an appropriate choice for the CCM model state $|\Phi\rangle$ is often 
a classical spin state, in which the most general 
situation is that each spin can point in
an arbitrary direction. In order to treat 
the Shastry-Sutherland model using the CCM, we choose 
the N\'eel state and the spiral state in Fig.(\ref{fig1}) to 
be our model states. We note that we do not choose the classical 
result for the pitch angle $\phi$ but we consider it rather 
as a free parameter in the CCM 
calculation. 

To treat each site equivalently we perform a rotation of the local axis of
the spins such that all spins in the reference state align in the same
direction, namely along the negative $z$ axis, such that we have 
$|\Phi\rangle \hspace{-3pt} = \hspace{-3pt}|\hspace{-3pt}\downarrow \rangle|\hspace{-3pt}\downarrow 
\rangle |\hspace{-3pt}\downarrow \rangle \ldots \,\,$. 
We define a set of multi-spin creation operators $C_I^+=s_r^+ \; , \;
s_r^+ s_l^+ \; , \; s_r^+ s_l^+ s_m^+\; , \; \ldots \;$.

The choice of the $C_I^+$ ensures that $\langle \Phi| C_I^+ = 0 =
C_I|\Phi\rangle$, where $C_I$ is the Hermitian adjoint of $C_I^+$. 

In order to make the spin $s_i$ to be aligned along the negative $z$ axis 
one has to perform a rotation of the respective spin by 
an appropriate angle $\delta_i$. 
This rotation is equivalent to the canonical transformations,  
\begin{eqnarray}
\label{eq3}  
s_i^x &=& \cos\delta_i s_i^x+\sin\delta_i s_i^z \nonumber \\
 s_i^y &=& s_i^y \nonumber\\
s_i^z &=& -\sin\delta_i s_i^x+\cos\delta_i s_i^z .
\end{eqnarray}
Using this transformation the Hamiltonian (\ref{eq1}) 
is then rewritten as
\begin{eqnarray}
\label{eq4}
&&H = J_1 \sum_{\langle i,j \rangle}^N \Big (\frac{1}{2}\sin\varphi_{i,j}[\hat{s}_i^+\hat{s}_j^z-
\hat{s}_i^z\hat{s}_j^++ \hat{s}_i^-\hat{s}_j^z-\hat{s}_i^z\hat{s}_j^-]\nonumber\\
&&+\cos\varphi_{i,j} \hat{s}_i^z\hat{s}_j^z 
+ \frac{1}{4}(\cos\varphi_{i,j}+1)[\hat{s}_i^+\hat{s}_j^- 
+ \hat{s}_i^-\hat{s}_j^+]\nonumber\\
&&+\frac{1}{4}(\cos\varphi_{i,j}-1)
[\hat{s}_i^+\hat{s}_j^+
+ \hat{s}_i^- \hat{s}_j^-]\Big )\nonumber\\
&&+\; J_2\sum_{\{i, k\}}^N\Big (\frac{1}{2}\sin\varphi_{i,k}
[\hat{s}_i^+\hat{s}_k^z-\hat{s}_i^z\hat{s}_k^+ + \hat{s}_i^-\hat{s}_k^z-\hat{s}_i^z\hat{s}_k^-] \nonumber\\
&&+\cos\varphi_{i,k} \hat{s}_i^z\hat{s}_k^z + \frac{1}{4}(\cos\varphi_{i,k}+1)[\hat{s}_i^+\hat{s}_k^-
+ \hat{s}_i^-\hat{s}_k^+]\nonumber\\
&&+\frac{1}{4}(\cos\varphi_{i,k}-1) 
[\hat{s}_i^ + \hat{s}_k^++\hat{s}_i^-\hat{s}_k^-]\Big ),
\end{eqnarray}
where the angles $\varphi_{i,j}\equiv\delta_j-\delta_i$, $\varphi_{i,k}\equiv\delta_k-\delta_i$ between
two nearest-neighbor and next-nearest-neighbor 
spins are $\varphi_{i,j}=\pi+\phi$,
$\varphi_{i,k}=-2\phi$,  respectively, and $s^{\pm}\equiv s^x \pm is^y$ are
spin raising and spin lowering operators.

The ket and bra GS's  $|\Psi\rangle$ and
$\langle\tilde{\Psi}|$ of $H$
are parametrised within the CCM as follows:
\begin{eqnarray}
\label{eq5}
H|\Psi\rangle = E|\Psi\rangle\,\, ; \qquad  \langle\tilde{\Psi}|H = E
\langle\tilde{\Psi}| \; \; ;\nonumber\\
|\Psi\rangle = e^S|\Phi\rangle\,\, ; \qquad S = \sum_{I \neq
0}\cal S_IC_I^+\; \; ; \nonumber\\
\langle\tilde{\Psi}| =  \langle\Phi|\tilde{S}e^{-S}\,\, ; 
\qquad \tilde{S} =
1 + \sum_{I \neq 0}\tilde{\cal S}_IC_I^- \;.
\end{eqnarray}
The correlation operators $S$ and $\tilde {S}$ contain the  correlation coefficients
${\cal S}_I$ and $\tilde {\cal S}_I$ which have to be determined.
Using the Schr\"odinger equation, $H|\Psi\ra=E|\Psi\ra$, we can now write
the GS energy as $E=\la\Phi|e^{-S}He^S|\Phi\ra$. After the 
notational rotation of the local axes of the quantum spins, 
the sublattice magnetization is given
by $ M = -1/N \sum_i^N \la\tilde\Psi|s_i^z|\Psi\ra$.

To find the ket-state and bra-state  correlation coefficients ${\cal S}_I$ and $\tilde {\cal S}_I$
we  require that the expectation value $\bar H=\la\tilde\Psi|H|\Psi\ra$ is
 a minimum with respect
to ${\cal S}_I$ and $\tilde {\cal S}_I$, such that the CCM ket-state and bra-state
equations are given by:
\begin{eqnarray}
\label{eq6}
\langle\Phi|C_I^-e^{-S}He^S|\Phi\rangle = 0 \qquad \forall I\neq
0\nonumber\\
\langle\Phi|\tilde{\cal S}e^{-S}[H, C_I^+]e^S|\Phi\rangle = 0 \qquad \forall
I\neq 0.\nonumber\\
\end{eqnarray}

The CCM formalism is exact if we take into account all possible multispin configurations in 
the correlation operators $S$ and $\tilde S$, which is, however, in general 
impossible for  a quantum many-body model. 
Hence, it is necessary to use approximation 
schemes in order to truncate the expansion 
of $S$ and $\tilde S$ in the Eqs.~(\ref{eq5})
in any practical calculation. 
The most common scheme is the LSUB$n$ scheme, 
where we include only $n$ or fewer correlated 
spins in all configurations 
(or lattice animals in the language of graph theory) 
which span a range of no more than $n$ adjacent (contiguous) 
lattice sites. 
\begin{table}[ht]
\caption{Number of fundamental GS configurations of the LSUB$n$
approximation for the Shastry-Sutherland model 
using the N\'eel state ($\phi=0$) and the spiral state ($\phi\neq 0$) 
as the  CCM reference state.}
\label{tab1}
\begin{tabular}{ccc}
\hline
\hline 
LSUB$n$ \,\,\,\,\,& N\'eel  state: $\phi=0$ \,\,\,\,\,&spiral state:  $\phi\neq 0$  \\
\colrule
2 &     1 &     12  \\
4 &    35 &    248 \\
6 &   794 &  6184  \\
8 &  20892& 166212 \\
\hline
\hline
\end{tabular}
\end{table}

To find all possible fundamental configurations which 
are different under the point and space group symmetries of both the lattice and the Hamiltonian, 
we use the lattice symmetries. The numbers of fundamental configurations may be 
further reduced 
by the use of additional conservation laws. For example, in the case of the N\'eel state $(\phi=0)$, 
the Hamiltonian of Eq.~(\ref{eq1}) commutes with the total uniform 
magnetization, 
$s_T^z=\sum_k s_k^z$ (the sum on $k$ runs over all lattice sites). the 
GS lies in the $s_T^z=0$ 
subspace, and hence we exclude configuration with an odd number of spins or with unequal numbers 
of spins on the two equivalent sublattices. For the spiral state we cannot apply this property because 
it is not an eigenstate of $s_T^z$. We calculate the fundamental configurations numerically, and 
the results of the numbers of LSUB$n$ configurations for $n \le 8$ are given in Table I. 
By using parallel computing we are able to solve the   20892 equations of
the CCM-LSUB8 approximation for the \Neel reference state. However, for the
spiral state the current limitations of computer power allow then solution
of the CCM equations up to LSUB6, only.  

Since the LSUB$n$ 
approximation becomes exact in the limit 
$n \to \infty$, it 
is useful to extrapolate the 'raw' LSUB$n$
results to the limit $n \to \infty$. 
Although an exact scaling theory for the LSUB$n$ results is not known, 
there is some empirical experience \cite{krueger00,zeng98,bishop00} 
how the physical quantities for antiferromagnetic 
spin models scale with $n$.
As stated above for the \Neel reference state 
we are able  to calculate
the GS energy $E$ and the sublattice magnetization $M$ within
LSUB$n$ up to $n=8$. 
In order to obtain more accurate results for the  
GS energy, we now employ a scaling law\cite{krueger00,bishop00}
in order to extrapolate our results in the limit 
$m \rightarrow \infty$, where  
\be \label{scal_e}
E(n) = a_0 + a_1\frac{1}{n^2} + a_2\left(\frac{1}{n^2}\right)^2 ~~.
\ee
We use CCM results for $n=4,6,8$  in order 
to carry out these extrapolations.\cite{bishop00} We find, however, that 
other scaling laws proposed in the literature 
yield very similar results for the energy. 
In the \Neel ordered phase we utilise \cite{bishop00} 
a scaling law with leading power $1/n$, i.e., 
\be \label{scal_m1}
M(n) = b_0 + b_1\frac{1}{n} + b_2\left(\frac{1}{n}\right)^2 ~~.
\ee
We find that this prescription again leads to reasonable 
results\cite{bishop00}. However, applying this scaling rule 
to systems showing an order-disorder transition at zero 
temperature this kind of scaling tends to overestimate 
the magnetic order and yields too large critical values for the 
exchange parameter driving the transition\cite{krueger00,rachid04}. 
The reason for that might         
consist in the change of the scaling near a critical point.
Hence in addition to the scaling rule (\ref{scal_m1}) we also 
use a leading 'power-law' scaling\cite{bishop00}, given by
\be \label{scal_m2}
 M(n)=c_0+c_1\left(\frac{1}{n}\right)^{c_2} ~~ .
\ee
The leading exponent $c_2$ is determined directly from 
the LSUB$n$ data.

\section{Results} \label{results}
We start with the discussion of the onset of the spiral phase in the quantum
model.
We calculate the GS energy as a function of $J_2$
using as reference state a spiral state as sketched in Fig. \ref{fig1}. As 
quantum fluctuations may lead to a ``quantum''
pitch angle that is different from the the classical case, we consider 
the pitch angle in the reference state as a free parameter. 
We then determine the ``quantum'' pitch angle $\phi_{qu}$ 
by minimizing $E_{{\rm LSUB}m}(\phi)$ with respect to $\phi$ 
in each order $n$. 
\begin{figure}[ht]
\vspace{0.3cm}
\psbild{\centerline{\epsfysize=6cm \epsfbox{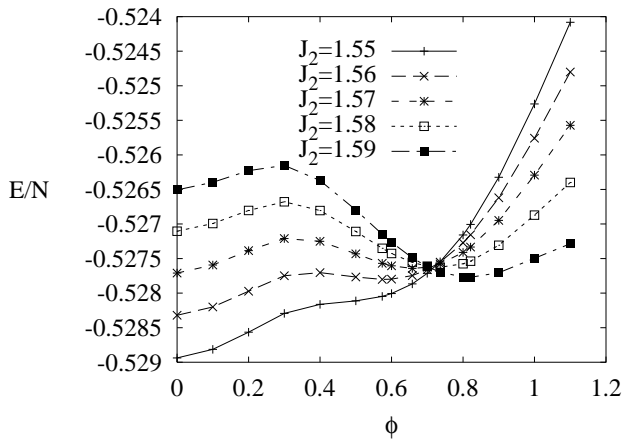}}}
\caption{Ground-state energy versus the pitch angle $\phi$ 
within CCM-LSUB4 approximation for different  values 
of $J_2$ in the range $1.55 \le J_2 \le 1.59 $.}
\label{fig2}
\end{figure}
\begin{figure}[ht]
\vspace{0.3cm}
\psbild{\centerline{\epsfysize=6cm \epsfbox{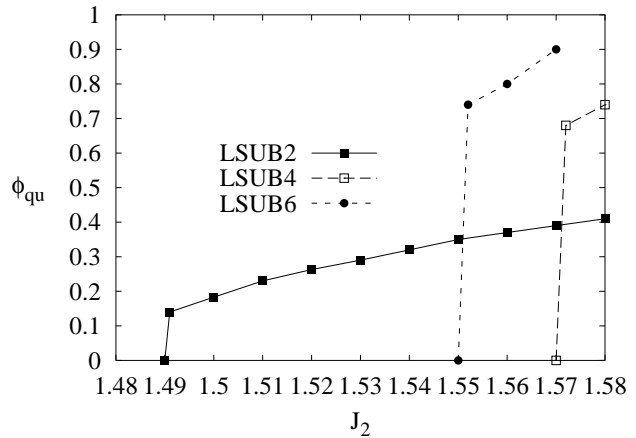}}}
\caption{The ``quantum'' pitch angle $\phi_{qu}$
 as a function of $J_2$ calculated within  
CCM-LSUB$n$ approximation with
$ n=2, 4, 6 $.}
\label{fig3}
\end{figure}
As for the classical model 
for small $J_2$ the energy $E_{{\rm LSUB}m}(\phi)$ has its 
minimum at $\phi_{qu}=0$, i.e., the quantum GS is 
the semi-classical collinear \Neel state.
Contrary to the classical case,   
this collinear quantum state can survive into the region $J_2 > J_1$,
 where classically it is already unstable.
This effect is known as {\it order from disorder} \cite{villain,shender} and
is widely observed in quantum spin systems, see, e.g., 
Refs. \onlinecite{bursill,krueger00}.  
For frustrating couplings $J_2 \gtrsim 1.5 J_2$ apart from the minimum at 
$\phi=0$ a second minimum at a finite $\phi > 0$ 
 emerges, which becomes the
global minimum for strong enough $J_2$. This scenario illustrated in Fig.
\ref{fig2} is  typical for a first-order transition, i.e., 
we find indications  that quantum 
fluctuations may change the nature of the phase transition between the 
the collinear \Neel phase to the noncollinear spiral phase 
from a  second-order classical 
transition to a first-order quantum transition.        
Note that a similar situation can be found in other frustrated spin systems
\cite{krueger00,krueger01}.
The ``quantum'' pitch angle $\phi_{qu}$, where 
$E_{{\rm LSUB}m}(\phi)$ has its global minimum, is shown in Fig. \ref{fig3}.
$\phi_{qu}$ shows a typical jump from $\phi_{qu}=0$ to a finite value.
Our data clearly indicate that the quantum noncollinear
spiral phase has lower energy
than the collinear phase only for strong frustration $J_2 \gtrsim 1.5J_1$. 

\begin{figure}[ht]
 \vspace{0.3cm}
\psbild{\centerline{\epsfysize=6cm \epsfbox{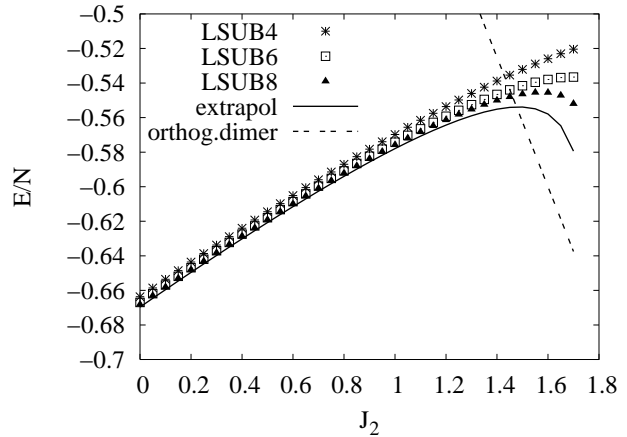}}}
\caption{The energy of (i) the collinear quantum ground state
as function of $J_2$ obtained by CCM-LSUB$n$ with 
$n=4, 6, 8$ and its extrapolated value to $n \to \infty$, see 
Eq.~(\ref{scal_e}), and (ii) 
of the orthogonal-dimer state.} 
\label{fig4}
\end{figure}

Next we compare the energy of the orthogonal-dimer state 
$|\Psi\rangle_{dimer}$ and the  energy  of the collinear quantum ground state
(i.e. the reference state $|\Phi_0\rangle$ is the \Neel
state), see Fig \ref{fig4}. We can postpone 
the discussion of the question whether 
that quantum ground state possesses \Neel LRO or not, since 
it is possible (starting from the  \Neel reference state) 
to calculate the energy up to high accuracy even in a parameter regime where
the \Neel order breaks down due to quantum fluctuations, i.e. 
for a magnetically disordered state, see e.g. Refs.
\onlinecite{squa_ref10,zeng98,krueger00,krueger01,Farnell04,rachid04}. 
Our results demonstrate, that the 
orthogonal-dimer state has lower energy than the collinear state for
$J_2 \gtrsim 1.477J_1$. $|\Psi\rangle_{dimer}$ 
remains the state of lowest energy also in the region where the
noncollinear spiral state has lower energy than the collinear phase.
We conclude that there is no intermediate spiral  phase 
in the quantum model. Our estimate
of the critical value   $J_2^{d}=1.477J_1$ 
where the transition to the orthogonal-dimer 
phase takes place is in good agreement with other results, cf.
Table 2 in Ref.
\onlinecite{Miyahara03}.

\begin{figure}[ht]
\vspace{0.3cm}
\psbild{\centerline{\epsfysize=6cm \epsfbox{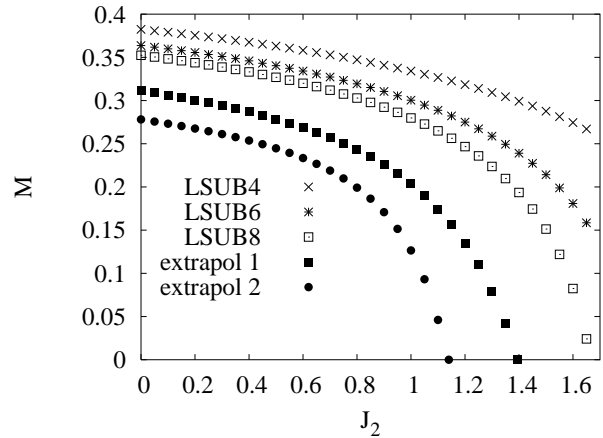}}}
\caption{Sublattice magnetization $M$ versus $J_2$ 
obtained by CCM-LSUB$n$ with 
$n=4, 6, 8$ and its extrapolated values to $n \to \infty$ 
using two different extrapolation schemes, namely according to
Eq.~(\ref{scal_m1}) (extrapol 1) and to Eq.~(\ref{scal_m2}) (extrapol 2).}
\label{fig5}
\end{figure}

So far we have 
discussed mainly the energy of competing GS phases.
The last question we would like to discuss is the question of the stability of
the \Neel LRO in the frustrated regime. For that we calculate the order
parameter (sublattice magnetization) $M$
within the LSUB$n$ approximation scheme up to $n=8$ 
and extrapolate to 
$n \to \infty$ using two variants of extrapolation 
as described in Sect.~\ref{ccm}. 
The results are shown in 
Fig.~\ref{fig5}.
The extrapolated data 
clearly demonstrate that the LRO vanishes before the 
orthogonal-dimer state becomes the GS. 
The transition from \Neel LRO to
magnetic disorder is of second order.
Hence we come to the second important statement that there exists 
an intermediate magnetically disordered phase.
Within the used CCM scheme starting from the \Neel reference state we are
not able to discuss the nature of the magnetically disordered state preceding
the orthogonal-dimer state. 
Though there are some first attempts to
develop a CCM formalism for magnetically disordered valence bond phases\cite{xian94}, 
a high level of
approximation is reached currently only starting with \Neel or spiral
reference states.

Obviously, the critical value where $J_2^{c}$ the \Neel LRO breaks down
depends on the used extrapolation formula. The extrapolation according to
Eq.~(\ref{scal_m1}) leads accurate results for $M$ in the unfrustrated
($J_2=0$) square-lattice limit and yields 
$J_2^{c} \sim 1.39 J_1$. As discussed in Sect.~\ref{ccm} this extrapolation
scheme tends to overestimate the region of magnetic LRO and indeed 
the value $J_2^{c}/J_1 = 1.39$ is significantly larger 
than the corresponding value calculated by series expansion, 
see Table 2 in Ref. \onlinecite{Miyahara03}. 
The extrapolation according to
Eq.~(\ref{scal_m2}) with a variable exponent $c_2$ is less accurate  
in the unfrustrated
limit but it seems ro be more appropriate to find the position of the critical
point $J_2^c$, since the scaling behavior might be changed at the critical
point. We get  $J_2^{c} \sim 1.14 J_1$ which fits well
to the corresponding value calculated by series expansion.
\section{Conclusions} \label{concl}
We have studied the GS phase diagram of the
spin half Shastry-Sutherland amtiferromagnet making use of 
high-order coupled cluster
calculations. Comparing the energies of competing \Neel, spiral and
orthogonal-dimer phases we can rule out the existence of a noncollinear
spiral phase. Considering the \Neel order parameter 
we find that the semi-classical \Neel long-range order 
disappears before the orthogonal-dimer phase sets in. Hence we conclude that 
the \Neel phase and the dimer phase are separated by a magnetically
disordered intermediate phase.

{\it   Acknowledgment:}
This work was supported by the DFG (project Ri615/12-1). The authors are
indebted to the Rechenzentrum of the University Magdeburg and in particular
to J. Schulenburg for assistance in numerical calculations.

\end{document}